\documentclass[a4paper]{spie}
\usepackage{graphicx}
\usepackage{amssymb}
\usepackage{hyperref}
\usepackage{subfig}
\usepackage{bchart}

\title{Infrared Interferometry and AGNs: Parsec-scale Disks and Dusty Outflows}

\author{Leonard Burtscher\supit{a}\footnote{\,\,\,burtscher@mpe.mpg.de}, Sebastian H\"onig\supit{b}, Walter Jaffe\supit{c}, Makoto Kishimoto\supit{d}, Noel Lopez-Gonzaga\supit{c}, Klaus Meisenheimer\supit{e}, Konrad R. W. Tristram\supit{f},
\skiplinehalf
\supit{a} Max-Planck-Institut f\"ur extraterrestrische Physik, Gie\ss enbachstr. 1, 85748 Garching, Germany\\
\supit{b} School of Physics \& Astronomy, University of Southampton, Southampton SO17 1BJ, UK\\
\supit{c} Sterrewacht Leiden, Niels Bohrweg 2, 2333 CA Leiden, The Netherlands\\
\supit{d} Kyoto Sangyo University, Kamigamo-motoyama, Kita-ku, Kyoto 603-8555, Japan\\
\supit{e} Max-Planck-Institut f\"ur Astronomie, K\"onigstuhl 17, 69117 Heidelberg, Germany\\
\supit{f} European Southern Observatory, Alonso de C\'ordova 3107, Vitacura, Santiago, Chile}

\begin{document}
\maketitle

\begin{abstract}
The ``torus'' is the central element of the most popular theory unifying various classes of AGNs, but it is usually described as "putative" because it has not been imaged yet. Since it is too small to be resolved with single-dish telescopes, one can only make indirect assumptions about its structure using models. Using infrared interferometry, however, we were able to resolve the circum-nuclear dust distributions for several nearby AGNs and achieved constraints on some further two dozen sources. We discovered circum-nuclear dust on parsec scales in all sources and, in two nearby sources, were able to dissect this dust into two distinct components. The compact component, a very thin disk, appears to be connected to the maser disk and the extended one, which is responsible for most of the mid-IR flux, is oriented perpendicularly to the circum-nuclear gas disks. What may come as a surprise when having in mind the standard unification cartoon actually connects well to observations on larger scales. Optically thin dust in the polar region, perhaps driven by a disk wind, could solve both the scale height problem of the torus and explain the missing anisotropy in the mid-IR -- X-ray relation.
\end{abstract}


\section{Introduction}
\label{sec:intro}

\begin{figure}
	\begin{center}
		\begin{tabular}{c}
			\includegraphics[width=10cm]{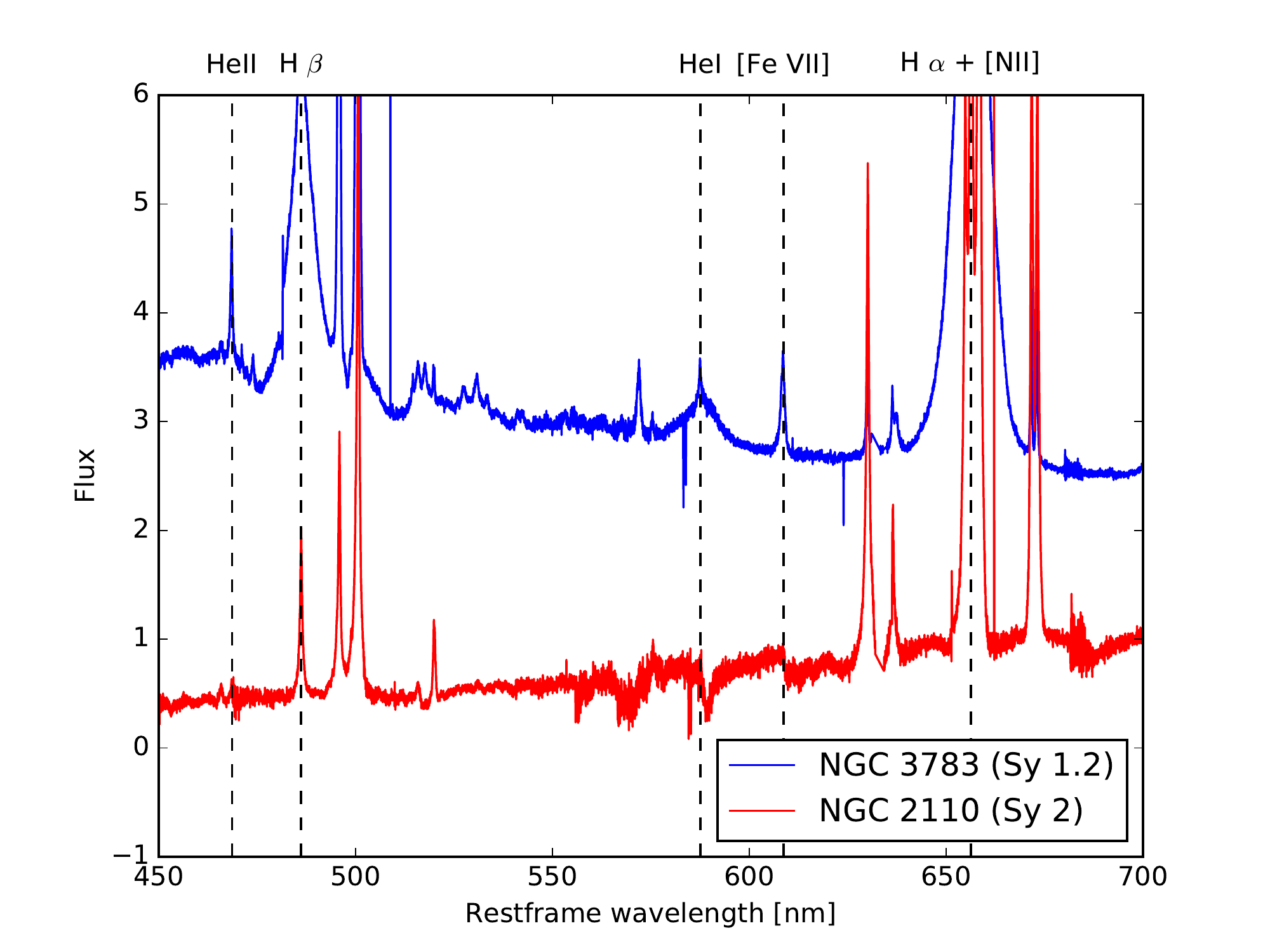}
		\end{tabular}
	\end{center}
	\caption{\label{fig:sy1_2} A spectrum of a Seyfert 1 galaxy (blue) and a Seyfert 2 galaxy (red) taken with the VLT/X-SHOOTER instrument as part of the LLAMA survey\cite{davies2015}. Both AGN types show narrow permitted and forbidden emission lines, but only the type 1 AGN also has broad permitted lines and a strong blue continuum. The sub-classification of Seyfert 1 galaxies (1.2, 1.5, ...) is based on emission line ratios.}
\end{figure} 

Optical spectra of active galactic nuclei (AGNs) can be classified into two distinct types depending on whether or not broad permitted emission lines are seen (Fig.~\ref{fig:sy1_2}). A Seyfert~1 (1.2, 1.5, 1.8, 1.9 or 1i) is an AGN where broad emission lines are detected (at variable strengths and wavelengths) while a Seyfert~2 galaxy does not show detectable broad emission lines. Since it has been observed that Seyfert~1 galaxies are more often seen face-on than Seyfert~2 galaxies, it has been speculated that obscuration by circum-nuclear dust may be responsible for the observed dichotomy\cite{keel1980}. Strong support has been given to the idea when Antonucci \& Miller showed in 1985\cite{antonucci1985} that broad lines are seen in the polarized light of the nearby prototypical Seyfert~2 galaxy NGC~1068. They interpreted these as scattered light from the intrinsic ``Seyfert~1'' nucleus of the source whose line of sight is blocked off by a geometrically and optically thick dusty disk, that was later\cite{krolik1986} dubbed ``torus''. This unification of Seyfert~1 and 2 galaxies by means of a dusty torus is a very popular idea and much research has been devoted to observe and model this dust distribution. We refer to Netzer 2015\cite{netzer2015} for a comprehensive recent review of AGN unification and will focus here on the contributions from infrared interferometry\footnote{More specifically, we will focus on long-baseline interferometry. We note that there are also two publications applying speckle interferometry on the nearby prototyptical AGN NGC~1068 in the H and K bands\cite{wittkowski1998,weigelt2004}.}.


The question of whether the ``unification'' model holds, has also implications for our understanding of the evolution of galaxies in general. A popular idea posits that mergers produce AGNs of high luminosity, i.e. quasars, which are thought to be first in an obscured and later in an unobscured phase\cite{sanders1988,hopkins2008}. Such an evolutionary scenario may not be applicable for lower luminosity, secularly evolving, systems but it is clear that the circum-nuclear gas and dust must play a key role in regulating the feeding of the AGN. A physical understanding of this region is therefore essential for grasping the effects AGN activity may have on the evolution of their host galaxies.

The ``torus'' picture and models based on that picture have been successful in explaining a number of observations including the detection of polarized broad lines\cite{ramosalmeida2016}, the collimation of ionization cones\cite{fischer2013}, its correspondence with the fraction of obscured sources\cite{maiolino1995} and the overall SED from the near- to far-infrared\cite{netzer2016}.


On the other hand, recent observations have also raised some questions. It is, for example, unclear whether ionization cones necessarily have to be collimated by the torus\cite{prieto2014}. Also, the detection of ``true'' type 2 objects, while rare, shows that some AGNs may not have a broad-line emitting region altogether\cite{shi2010} and modeling of the integrated ``torus'' SEDs (and later also of interferometric visibilities) have shown that the ``tori'' of type 2 Seyfert galaxies intrinsically have a higher covering factor and a larger number of dust-containing gas clumps\cite{ramosalmeida2011,elitzur2012,lopezgonzaga2016b}. These results already rule out the strictest (and astrophysically questionable) versions of ``unification'' requiring an identical ``torus'' in every AGN.

It it also challenging to explain the small, if any, anisotropy in the mid-IR emission from the ``torus'' at a given X-ray luminosity: Type 1 and type 2 AGNs essentially follow the same mid-IR--X-ray relation from low to high luminosities\cite{lutz2004,gandhi2009,asmus2011,asmus2015}. Apart from these observational complications, it is also unclear theoretically how a geometrically thick torus should exist dynamically in a disk configuration. If the dust were smoothly distributed and thermal random motion provided support against gravity, the large temperatures involved would immediately destroy the dust. The dust therefore has to be organized in clumps\cite{krolik1988}. What stabilizes these clumps against the surrounding medium or if they are stable at all is still an open question.

Much of the uncertainty about the geometry and dynamics of the ``torus'' comes from the fact that the circum-nuclear dust in AGNs is usually unresolved in single-dish high-resolution images\cite{asmus2014} -- a deficiency that infrared interferometry has partly solved in the last decade.

\section{Interferometric AGN observations}
\label{sec:obs}

\begin{table}[htbp]
   \centering
   \begin{tabular}{llll} 
      \hline
      Reference & Interferometer & \# & summary of result \\
      \hline
      Swain et al. 2003\cite{swain2003} & KI & 1 & Marginally resolved emission in NGC~4151\\
      Wittkowski et al. 2004\cite{wittkowski2004} & VINCI & 1 & \begin{tabular}{@{}l@{}}Low near-IR visibility for NGC~1068 argues for two-component \\ model\end{tabular}\\
      Jaffe et al. 2004\cite{jaffe2004} & MIDI & 1 & Resolved two components of warm and hot dust in NGC~1068\\
      Poncelet et al. 2006\cite{poncelet2006} & MIDI & 1 & Re-analysis of the Jaffe et al. 2004 MIDI data, find no hot dust\\
      Meisenheimer et al. 2007\cite{meisenheimer2007} & MIDI & 1 & Nucleus of Centaurus~A: a dusty disk and synchrotron emission\\
      Tristram et al. 2007\cite{tristram2007b} & MIDI & 1 & \begin{tabular}{@{}l@{}}Two-component structure of nuclear dust in Circinus, disk \\ component is warm and co-aligned with maser disk\end{tabular}\\
      Beckert et al. 2008\cite{beckert2008} & MIDI & 1 & Nuclear dust in NGC~3783 consistent with clumpy torus model\\
      Kishimoto et al. 2009a\cite{kishimoto2009} & MIDI \& KI & 4 & Evidence for a common radial structure in AGN tori\\
      Raban et al. 2009\cite{raban2009} & MIDI & 1 & \begin{tabular}{@{}l@{}}Two-component structure of nuclear dust in NGC~1068, disk \\ component is hot and co-aligned with maser disk\end{tabular}\\
      Tristram et al. 2009\cite{tristram2009} & MIDI & 8 & \begin{tabular}{@{}l@{}}Mid-IR sizes roughly scale with $\sqrt{L}$, no clear distinction \\ between type 1 and type 2 sources\end{tabular}\\
      Burtscher et al. 2009\cite{burtscher2009} & MIDI & 1 & \begin{tabular}{@{}l@{}}The nuclear dust in the Seyfert 1 galaxy NGC~4151 has similar \\ properties as in Seyfert 2 galaxies\end{tabular}\\
      Kishimoto et al. 2009b\cite{kishimoto2009b} & KI & 4 & \begin{tabular}{@{}l@{}}Interferometrically derived near-IR radii are slightly larger than\\ reverberation-based radii and therefore likely probing the\\ sublimation radius\end{tabular}\\
      Pott et al. 2010\cite{pott2010} & KI & 1 & \begin{tabular}{@{}l@{}}No change in near-IR size of circum-nuclear dust in NGC~4151 \\ despite variable luminosity\end{tabular}\\
      Burtscher et al. 2010\cite{burtscher2010} & MIDI & 1 & New mid-IR visibilities of Cen~A do not fit well to a dust disk\\
      Kishimoto et al. 2011a\cite{kishimoto2011} & KI & 8 & Sublimation radius scales with $\sqrt{L}$\\
      \begin{tabular}{@{}l@{}}Tristram \&\\ Schartmann 2011\end{tabular}\cite{tristram2011} & MIDI & 10 & \begin{tabular}{@{}l@{}}Differences in mid-IR sizes between type~1 and type~2 sources,\\ expected from models, are not seen observationally.\end{tabular}\\
      Kishimoto et al. 2011b\cite{kishimoto2011b} & MIDI & 6 & Half-light radius in the mid-IR independent of luminosity\\
      Weigelt et al. 2012\cite{weigelt2012} & AMBER & 1 & Marginally resolved near-IR emission in NGC~3783\\
      H\"onig et al. 2012\cite{hoenig2012} & MIDI & 1 & \begin{tabular}{@{}l@{}}Majority of mid-IR emission originates from optically thin dust\\ in the polar region in NGC~424 and is part of the outflow\end{tabular}\\
      H\"onig et al. 2013\cite{hoenig2013} & MIDI & 1 & Detection of dust in the polar region of NGC~3783\\
      Burtscher et al. 2013\cite{burtscher2013} & MIDI & 23 & \begin{tabular}{@{}l@{}}MIDI AGN Large Programme results: half-light radius in the\\ mid-IR scales with luminosity, but with large scatter;\\ tori show a large diversity in intrinsic structure\end{tabular}\\
      Kishimoto et al. 2013\cite{kishimoto2013} & KI & 7 & Evidence for a receding dust sublimation region in NGC~4151\\
      Tristram et al. 2014\cite{tristram2014} & MIDI & 1 & \begin{tabular}{@{}l@{}}Updated model for the Circinus~galaxy including data from \\ shorter AT baselines; two-component structure confirmed with \\ larger structure in the polar direction; SED model predicts \\ sub-mm flux precisely as measured with ALMA;\\ no evidence for large amounts of cold gas\end{tabular}\\
      Lopez-Gonzaga et al. 2014\cite{lopezgonzaga2014} & MIDI & 1 & \begin{tabular}{@{}l@{}}Updated model for NGC~1068 including data from shorter\\ AT baselines; two component structure confirmed with \\ larger structure in the polar direction\end{tabular}\\
      Lopez-Gonzaga et al. 2016\cite{lopezgonzaga2016b} & MIDI & 23 & \begin{tabular}{@{}l@{}}Modeling shows that the observed $(u,v)$ coverages only allow\\ to detect elongations in 7/23 sources.\\ 5/7 are found to be significantly elongated, all in polar direction.\end{tabular}\\
   \hline
   \end{tabular}
   \caption{Summary of all publications using long-baseline interferometry for studying AGNs (either data publication or new analysis). The third column gives the number of sources involved in the particular study.}
   \label{tab:results}
\end{table}

Since the first interferometric observation of AGNs in the infrared\cite{swain2003} with the Keck Interferometer\cite{vasisht2003}, such observations have also been successfully performed with three interferometers at the VLTI: VINCI\cite{kervella2000,kervella2003}, MIDI\cite{leinert2003} and AMBER\cite{petrov2007}. These interferometers have in common that they are sensitive at near- or mid-infrared wavelengths, are assisted by AO systems and that they are fed by large, 8-10m class telescopes.


Observing AGNs by means of infrared interferometry is ``hard'' because AGNs are not very bright compared to more typical targets of stellar interferometers. For example, NGC~1068 and the Circinus~galaxy, the brightest AGNs in the southern sky, offer a flux density of about 10~Jy at 12~$\mu$m, comparable to the faint end of objects usually studied with this method\cite{leinert2004,boley2013}. All other targets are at least about ten times fainter. The situation is even more severe in the near-infrared where the atmosphere changes much more rapidly, requiring much shorter fringe exposures and fast AO performances. In addition, it turned out that the angular size of the structure in the near-infrared was found to be quite small at least for type 1 AGNs where AO works well with the optically bright nucleus. Thus, with the currently available baseline lengths, the near-infrared structure is only marginally resolved, except for NGC~1068 which is the only type 2 AGN successfully observed with near-infrared interferometry so far.

\begin{figure}
	\begin{center}
		\begin{tabular}{c}
			\includegraphics[width=10cm]{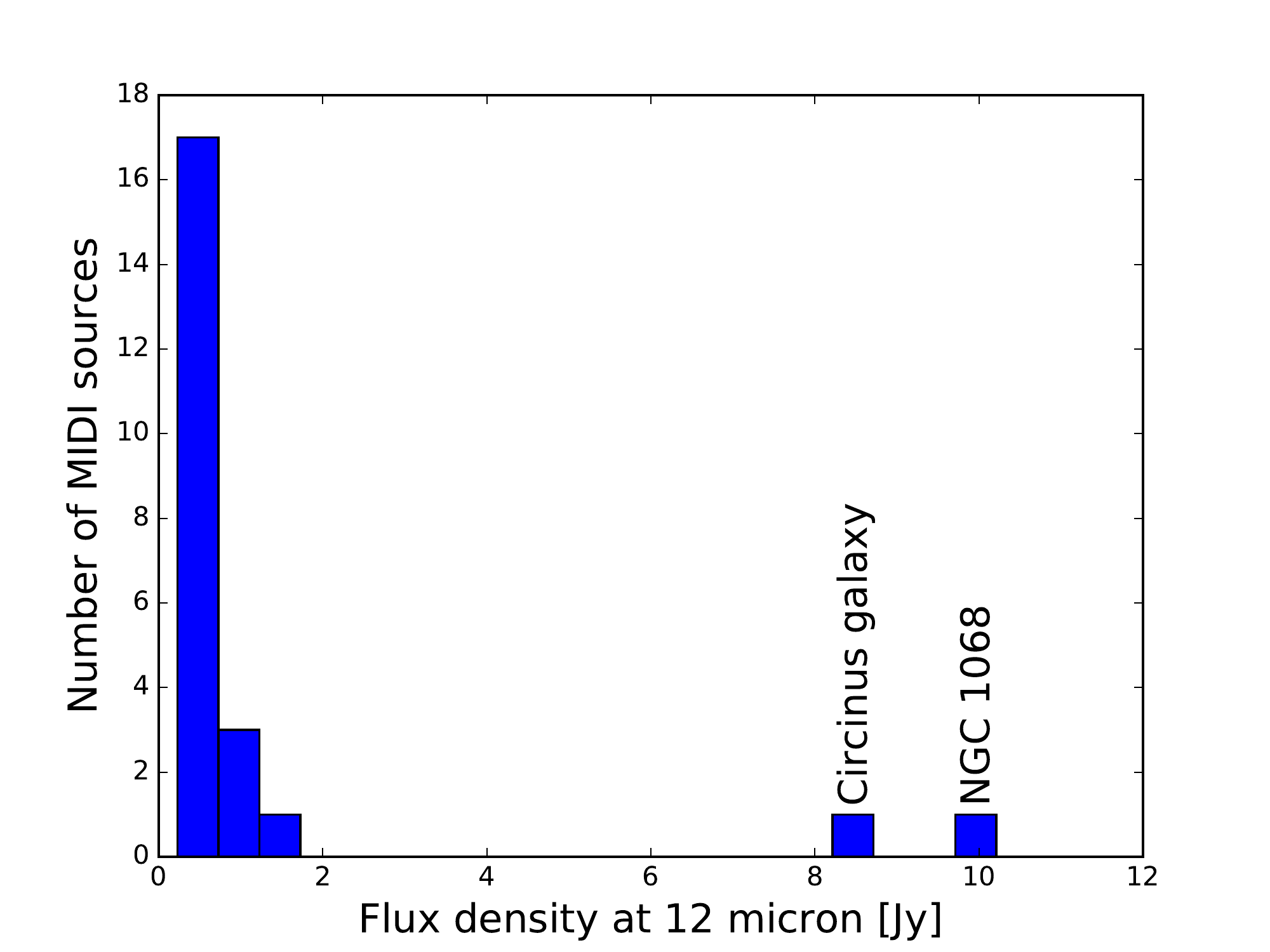}
		\end{tabular}
	\end{center}
	\caption{\label{fig:midi_lp_flux_histogram} Histogram of the total 12 $\mu$m fluxes of AGNs successfully observed with MIDI. Compared to the two moderately bright sources, NGC~1068 and the Circinus~galaxy, all other sources are about a factor of ten fainter.}
\end{figure}


Most AGNs were therefore studied in the mid-infrared with MIDI at the VLTI, which was the only mid-IR interferometer so far that was sensitive enough to observe AGNs. There are, however, a number of technical challenges for studying such faint sources in the thermal infrared, i.e. in the atmospheric window at 8--13 $\mu$m. One of the difficulties in these observations was the fact that the standard visibility calibration is a very noisy quantity for faint mid-infrared sources since it involves comparing the correlated with the uncorrelated flux of the target. The latter is very difficult to measure due to the intense and fast varying background in the mid-infrared, aggravated by the fact that MIDI observes the sky after about 20 reflections on warm mirrors along the VLTI beam train. This can be circumvented by comparing directly the correlated fluxes, which are less sensitive to this noise since the noise is largely uncorrelated. But when comparing fluxes directly, it is necessary to know the stability of the atmospheric transfer function as well as the total flux of the target from a separate, but ideally close-in-time, observation. An additional challenge for the group-delay fringe tracking mechanism of MIDI is that coherence losses start to become a significant problem for a signal-to-noise-ratio (SNR) lower than about 10. This problem was mitigated by measuring the coherence loss on ``fake weak'' sources (i.e. bright sources mixed with noise) and deriving a correction function to apply to science data. Both of these calibration steps have been successfully employed for the faint AGNs observed as part of the MIDI AGN Large Programme\cite{burtscher2012,burtscher2013}, but see also an alternative approach for faint-object calibration with MIDI\cite{kishimoto2011b}.

\section{Results and Discussion}
\label{sec:results}

Until now, about 40 AGNs have been studied at near- and mid-infrared wavelengths with interferometry, leading to a total of 24 refereed papers. A short summary of all these publications is given in Tab.~\ref{tab:results} and here below we highlight a few of them and give a more general summary.

\subsection{Near-infrared: The dust sublimation radius}
Near-infrared interferometry, mostly at the Keck Interferometer, has shown that the majority of the nuclear near-infrared emission originates from thermal radiation of dust at or very close to the sublimation radius\cite{kishimoto2009b} and that this radius scales with $\sqrt{L}$, where $L$ is the luminosity of the AGN\cite{kishimoto2011}. These sizes are slightly larger than those derived from earlier observations of this region using dust reverberation mapping\cite{suganuma2006}, but the discrepancy is likely due to the well-known difference between flux- and response-weighted radii, respectively\cite{kishimoto2009b}. A similar effect is seen when comparing the H$\beta$ radii of the broad line region derived from reverberation mapping\cite{bentz2009b} and from comparing integrated spectroscopy with photoionization calculations (Schnorr-M\"uller et al., submitted).

\subsection{Mid-infrared: A disk and a wind}
\subsubsection{The best studied object: The Circinus galaxy}
\begin{figure}
	\begin{center}
		\begin{tabular}{c}
			\includegraphics[width=10cm]{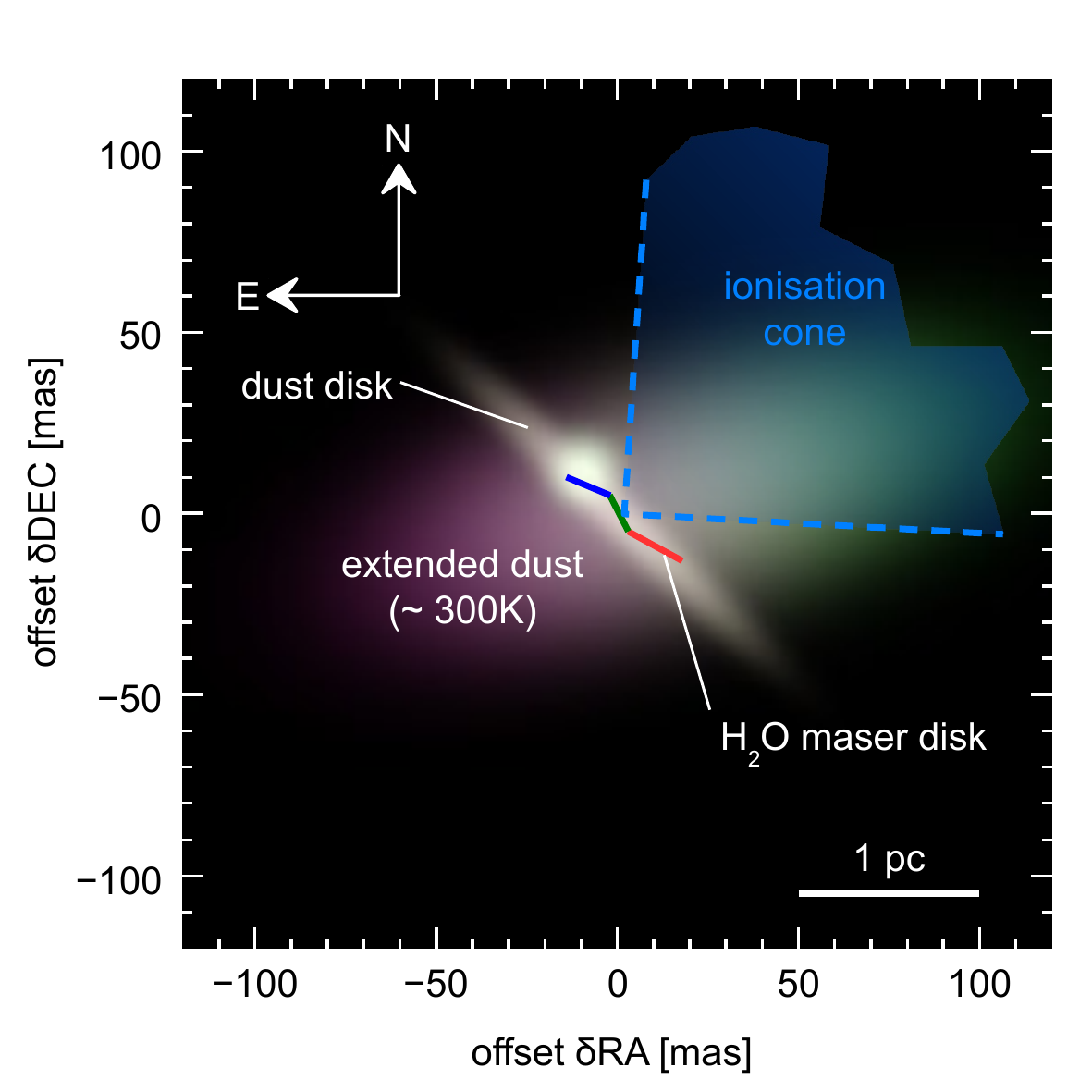}
		\end{tabular}
	\end{center}
	\caption{\label{fig:circinus} Model image for the VLTI/MIDI visibilities of the Circinus~galaxy from Tristram et al. 2014\cite{tristram2014}. The nuclear mid-IR flux is decomposed into three components: a 2-pc long, polar-elongated component with a large Silicate depth gradient (purple-green) contains most of the flux; a disk component (FWHM $\sim$ 1 pc) with a large axis ratio at the same position angle as the warped water maser disk (red--green-blue lines) and some unresolved flux (the bright spot at the blue end of the maser disk).}
\end{figure}

The Circinus~galaxy is the ideal target for interferometric observations from Cerro Paranal. It is relatively bright in the mid-IR, highly resolved and, due to its southern declination, offers a nearly circularly symmetric $(u,v)$ coverage allowing to probe many different position angles at various spatial scales. Thanks to its highly resolved structure and its relatively large signal-to-noise ratio, the nuclear mid-infrared flux of the Circinus~galaxy has been dissected into three distinct components (see Fig.~\ref{fig:circinus})\cite{tristram2007,tristram2014}:

\begin{enumerate}
	\item An elongated component of about 2~pc in size, containing the majority ($\sim 80\%$) of the flux. It seems to trace the southern edge of the near-side (north-west) ionization cone which also shows an enhanced flux in optical images\cite{wilson2000}. The observed wavelength-differential (or chromatic) phases observed for the Circinus~galaxy can be explained by introducing a strong gradient in the Silicate absorption feature of this component. This is consistent with the orientation of the host galaxy (i.e. the north-west is the near side with lower extinction).
	\item A highly elongated component of about 1 pc full-width at half-maximum (FWHM). It is oriented roughly perpendicularly to the large component and in the same direction as the emission traced by 22~GHz water masers. We therefore identify it with the highly inclined and warped maser disk, but the proof of this co-location will only come from the next generation of interferometers.
	\item An unresolved component containing $\sim 5\%$ of the total flux, potentially caused by radiative transfer effects in the dusty warped maser disk (Jud et al., submitted).
\end{enumerate}

A detailed model image and flux decomposition has also been achieved for the only other bright AGN target for MIDI, the prototypical Seyfert~2 galaxy NGC~1068\cite{jaffe2004,raban2009,lopezgonzaga2014}, but it suffers from a very unfortunate $(u,v)$ coverage due to its declination near $0^{\circ}$. The usage of the VLTI Sub-Array (VISA), i.e. the smaller Auxiliary Telescopes (ATs), has allowed us to better characterize the extended component in this galaxy and determine its orientation in the polar direction as well. Attempts to use the re-locatable AT baselines to improve the  $(u,v)$ coverage also for south-eastern baselines have failed, however, since correlated fluxes in this direction were too low for MIDI to rely on its internal group-delay tracking. Using PRIMA as a fringe tracker for MIDI\cite{mueller2010,pott2012} did not improve the situation since PRIMA was not sensitive enough to track fringes at this $(u,v)$ position either\footnote{The combination of PRIMA+MIDI worked well for stars, but not for AGNs whose typical nuclear $K-N$ colors are very red: 4--7 magnitudes (mag) for type 1 AGNs and 6--10 mag for type 2 AGNs\cite{burtscher2015}.}.

\subsubsection{Extending the sample}

Extending the sample of AGNs observed with MIDI proved difficult because the next brightest sources are about ten times fainter than the two sources mentioned above (see Fig.~\ref{fig:midi_lp_flux_histogram}). With a typical noise level in the correlated flux of about 0.05~Jy, these sources were observed at signal-to-noise ratios lower than 10 which caused artifacts in the group delay tracking (see above) that needed to be corrected.

Apart from the low fluxes, the sources of this extended sample were also much more compact than expected. For the inner edge of the dust distribution, $r_{\rm in}$, the scaling of the radius with luminosity $L$ is well established (see above and Tab.~\ref{tab:results}) and follows a $\sqrt{L}$ law over about four orders of magnitude in UV luminosity \cite{kishimoto2009b}. With the observed flux $F \propto L/D^2$ ($D$ being the distance to the source) and the observed size (angle) of the structure $\theta_{\rm in} \propto r_{\rm in}/D$, the angle the source subtends scales $\propto \sqrt{F}$. If all sources had the same intrinsic structure, we would expect the radius measured in the mid-infrared $\Theta_{\rm MIR}$ to scale similarly $\propto \sqrt{F_{\rm MIR}}$ where $F_{\rm MIR}$ is the observed flux in the mid-infrared. The fainter sources observed with MIDI do, however, not follow such a simple scaling relation. This issue has caused some controversy and led to papers claiming\cite{tristram2011} that the mid-IR size scales with $\sqrt{L}$, much flatter than this\cite{kishimoto2011b} and most recently and with the largest number of sources: that a relation exists but with a large intrinsic scatter\cite{burtscher2013}.

\begin{figure*}
	\centering
	\includegraphics[width=12cm]{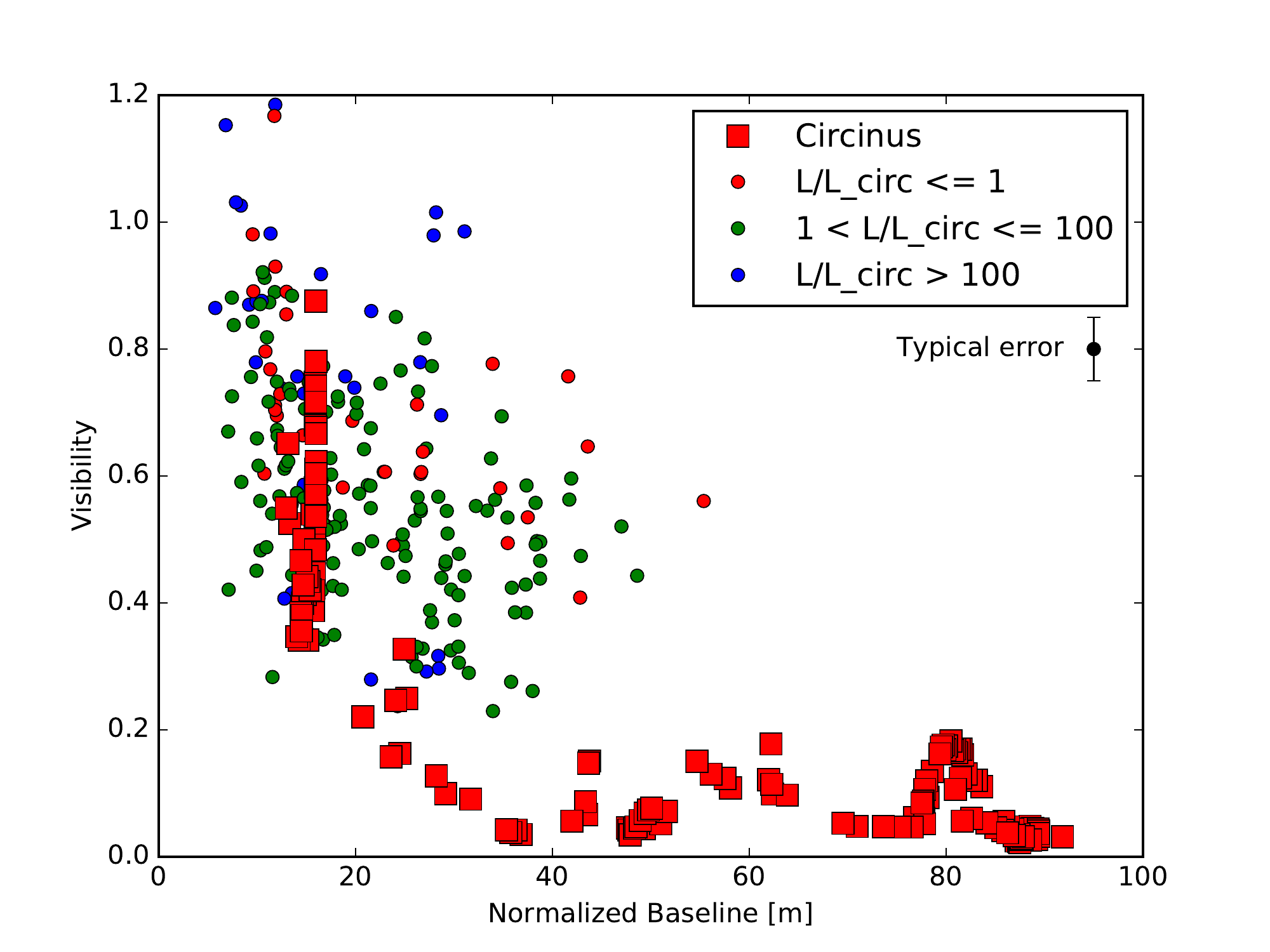}
	\caption{\label{fig:LP_rescale} Visibilities of the relatively bright Circinus~galaxy compared with the sources of the MIDI AGN Large Programme\cite{burtscher2013}. The baselines are normalized to the mid-IR flux of the Circinus~galaxy under the expectation that the mid-IR size scales $\propto \sqrt{F}$. The color encodes the mid-IR luminosity of the sources, compared to the mid-IR luminosity of the Circinus~galaxy $L_{\rm circ}$. A typical error bar is shown below the legend. It is unclear what causes the large intrinsic diversity of dust structures.}
\end{figure*}

\begin{figure}
	\begin{center}
		\begin{tabular}{c}
			\includegraphics[width=10cm]{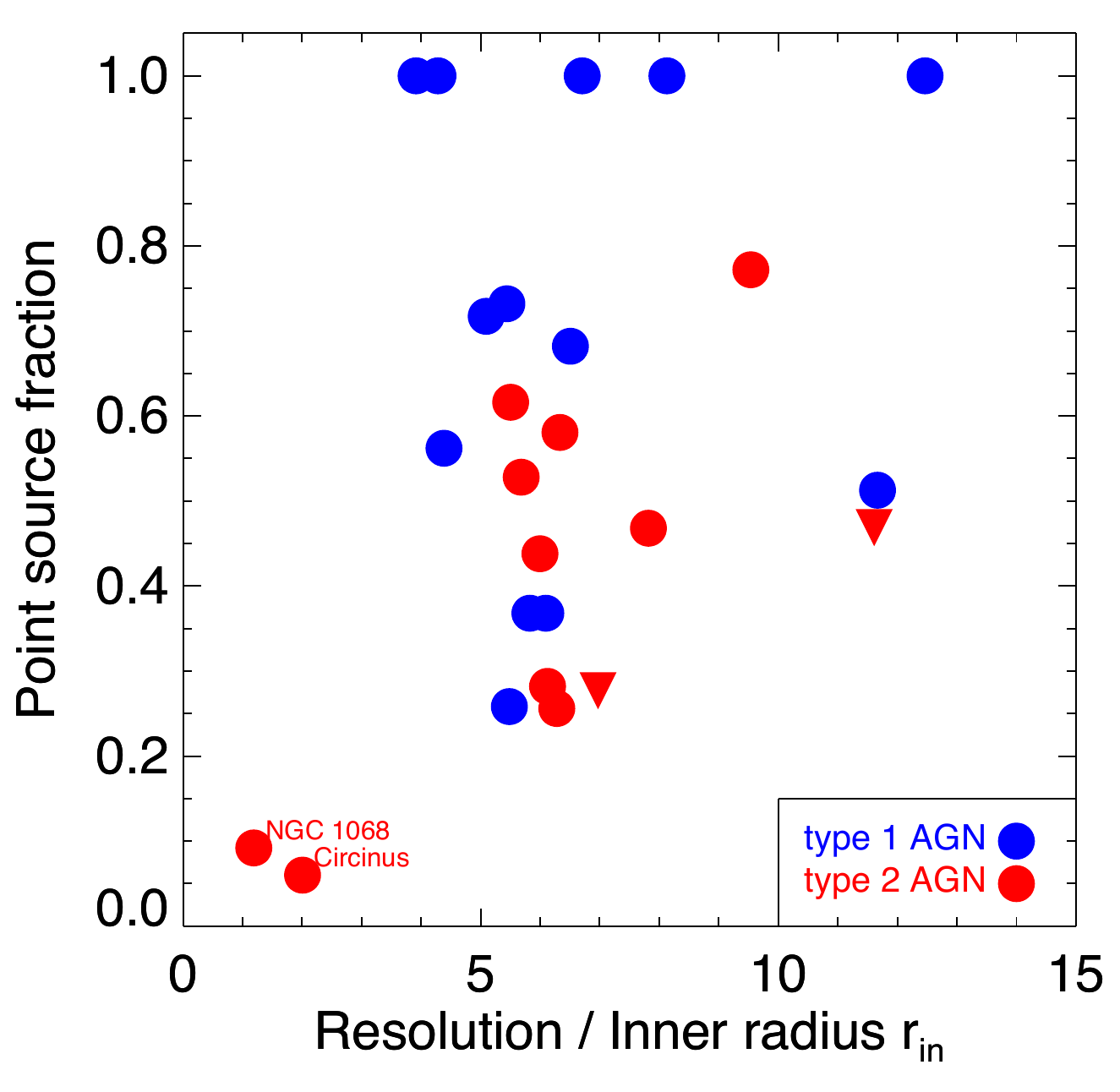}
		\end{tabular}
	\end{center}
	\caption{\label{fig:fp} Lowest visibilities (``point source fractions'') measured for the sample of 23 AGNs presented in Burtscher et al. 2013\cite{burtscher2013}. These visibilities are shown against the intrinsic resolution, i.e. the resolution in terms of the sublimation radius of dust. While the intrinsic resolution for the faint sources is comparable, they show a large diversity in unresolved flux. Type 1 AGNs are plotted in blue, type 2 AGNs in red.}
\end{figure}

One way to visualize the observed diversity in {\em intrinsic} dust structure is to plot the observed visibilities\footnote{Note that we actually observe and model correlated fluxes directly for reasons of better calibratability (see above). The visibilities here are generated from the ratio of the correlated flux and the total flux, where the latter has been observed separately with a single-dish mid-IR instrument, mostly VLT/VISIR.} as a function baseline normalized to Circinus, see Fig.~\ref{fig:LP_rescale}. This normalization takes into account the expected scaling with flux, so that the normalized baseline is given by the observed baseline $\times \sqrt{F/F_{\rm Circinus}}$. As can be seen, the fainter sources (plotted as small circles) are less well resolved than expected from scaling the Circinus visibilities (large red squares) according to the sublimation radius of dust. Alternatively, one can also look only at the lowest visibility as a function of intrinsic resolution (i.e. resolution scaled to the sublimation radius)\cite{burtscher2013}, see Fig.~\ref{fig:fp}. This figure shows that the fraction of unresolved source (the ``point source fraction'') is largely independent of resolution. What drives the intrinsic diversity in the dusty environment of the AGN is not clear yet. The luminosity seems to have only a small effect (see Fig.~\ref{fig:LP_rescale}). For the inner radius, a possible scaling with the radio-to-optical luminosity has been explored and a tentative correlation has been found\cite{kishimoto2011} indicating that the structure of the circum-nuclear dust could be related to the accretion mechanism on sub-parsec scales.

The reason for the larger scatter in the (normalized) mid-infrared sizes compared to the near-infrared is that the sublimation temperature sets a physical inner boundary while a unique temperature-radius relation does not exist for the warm dust. This is consistent with clumpy torus models that predict a very shallow dependence of brightness temperature with radius (e.g. Fig.~10 in Tristram et al. 2007\cite{tristram2007b}).



\subsection{Dust in the polar region}

Perhaps the most spectacular result from AGN interferometry is the detection of dust in the polar region of several nearby AGNs\cite{hoenig2012,hoenig2013,tristram2014,lopezgonzaga2014}. Initially this was interpreted as the hot inner funnel of a clumpy toroidal distribution of dust (some torus models predict `X' shaped or polar emission from azimuthal, ``torus''-like, dust structures)\cite{tristram2014}. This interpretation suggested that radiative transfer effects played a large role in the apparent distribution of mid-infrared flux. A new interpretation of the polar emission as optically-thin emission from the sheath of the outflowing ionization cone\cite{hoenig2012,hoenig2013} was supported by a high-resolution SED compiled for the Circinus~galaxy. A high-resolution ALMA continuum measurement\cite{hagiwara2013} agreed extremely well with the flux predicted from modeling the warm dusty structure. This implied that the sub-mm flux is the Rayleigh-Jeans tail of the warm dusty structure and there is not any more cold dust that is hidden in the mid-IR.\cite{tristram2014}

A study assessing the frequency of polar dust emission examined the detectability of elongated structures given the very limited $(u,v)$ coverages of the observations in combination with observed noise levels and the level of unresolved flux \cite{lopezgonzaga2016a}. It was found that, given these observational limitations, only in seven out of 23 sources can a possible elongation (axis ratio and position angle) be constrained. And among these seven sources there are five that actually have elongated emission, all of them oriented much closer to the polar direction than orthogonal to it. This is consistent with observations on larger scales that typically find dust in the narrow-line region as well\cite{cameron1993,schweitzer2008,asmus2016}. However, single dish imaging also always finds a strong nuclear point source which only interferometry could resolve to see that most of the circum-nuclear parsec-scale dust emission originates from the outflow region, rather than from a ``torus''.



\section{Conclusions}
Infrared interferometry, and especially mid-infrared interferometry with MIDI at the VLTI, has been the key technology to dissect the otherwise unresolved nuclear dust structure in AGNs. It has become clear that the majority of the dust emission, and apparently also of the dust mass, is located in or near the outflow cone and possibly constitutes a dusty sheath around the ionizing cone, perhaps contributing to its collimation. This naturally solves the long-standing scale-height problem of the ``torus''. If the dust emission in the polar region is optically thin in the mid-IR as the good agreement of the sub-mm continuum with the extrapolated mid-IR model suggests, this would also solve the problem of essentially isotropic mid-IR emission as seen in the mid-IR -- X-ray correlation. The collimation of the ionizing radiation would then happen at the base of the dusty outflow (requiring only relatively little obscuration). The same structure could then also be responsible for the observed dichotomy in Seyfert spectra. Alternatively, the thin disk may provide the obscuration required to block our view onto the broad line region in Seyfert~1 galaxies. Its scale height as measured in the mid-IR and seen in the maser disks may not be sufficient, although the warp in the maser disk may help\cite{lawrence2010}. In the inner parts, this disk may have a puffed-up inner rim as suggested by the relatively large near-IR covering factors\cite{hoenig2013} and thus provide the opening angle required to explain the fraction of obscured sources.

It seems not exaggerated to blame mid-IR interferometry (and VLTI/MIDI in particular) for ``killing'' the postulated AGN torus by dissecting it into a thin disk and a polar-elongated outflow, none of which are represented in the classical AGN torus cartoon.

\section{Outlook}
The two main interferometric instruments used for observing AGNs, the Keck Interferometer in the near-infrared and VLTI/MIDI for the mid-infrared, are no longer operational, but a new generation of interferometers is arriving at the VLTI. GRAVITY\cite{eisenhauer2014} is already producing first science results and will be used to study AGNs in the near-infrared soon. With its superb sensitivity and fringe-tracking capabilities, it will reach of order 20 AGNs. The scientifically perhaps most thrilling prospect for AGN science is to use spectro-astrometry to measure the size and geometry of the broad line region\footnote{We note early attempts of using spectro-astrometry with VLTI/AMBER by R. Petrov et al.\cite{petrov2012}}. It will also image the inner rim of dust, measuring its size and orientation. In combination with infrared continuum reverberation mapping the measured angular size of the sublimation radius provides an independent distance estimate for nearby AGNs\cite{hoenig2014}.

The next generation mid-IR interferometer MATISSE\cite{lopez2014}, scheduled to arrive on Paranal in about two years, will offer four-telescope imaging interferometry in the L, M and N bands. With real images and higher resolution (shorter wavelengths), it will provide more reliable estimates for the geometry of the circumnuclear dust, possibly revealing what drives the dichotomy in its intrinsic properties.

In order to observe a significant number of AGNs, however, MATISSE will require a sensitive external fringe tracker, such as provided by the GRAVITY instrument. These authors therefore welcome very much ESO's efforts to make GRAVITY available as a fringe tracker for MATISSE (see the ``GRA4MAT'' project in these proceedings).

\section{Acknowledgements}
The authors would like to thank the VLTI teams at ESO for tremendous support over many years of observations. L.B. is supported by a DFG grant within the SPP 1573 ``Physics of the interstellar medium''

\bibliography{SPIE_2016_LB.bbl}   
\bibliographystyle{spiebib}   

\end{document}